# On possibilities of application of Miller formula for determination of parameters of Micropixel Avalanche Photodiodes[1]

Z. Sadygov[2]**, Kh. Abdullaev, G. Akhmedov, F. Akhmedov, R. Mukhtarov, A. Sadygov, A. Titov, V. Zhezher

Joint Institute for Nuclear Research, 141980, Dubna, Moscow Reg., Russia

Miller formula modified to take into account voltage drop on serial resistor of an avalanche photodiode is considered. It is proven by experimental data that modified Miller formula can describe operation of both regular and micropixel avalanche photodiodes with good enough precision. The Miller parameter must be determined from a static voltage-current curve of the photocurrent in an avalanche photodiode. It is shown that operation parameters of the devices can be determined using a linear extrapolation of the voltage-current curve for both regular avalanche photodiode and the one operating in Geiger mode.

## 1. Introduction

Micropixel avalanche photodiodes (MAPD) also known as silicon photoelectron multipliers (SiPM) are widely used and planned for use in various scientific and industrial applications [1, 2, 3]. Since MAPD design and operation principles are different from those of traditional avalanche photodiodes (APD) it is necessary to develop new techniques for investigation of their parameters. MAPD sensitive area consists of independent p–n-junctions (micropixels) usually no more than *100 µm × 100 µm* in dimensions that are connected to a common bus via individual microresistors. Each micropixel can operate above breakdown voltage in so called Geiger mode and its gain can exceed $10^6$. For this reason not all techniques of measurement of APD parameters can be applied to MAPD.

Presented work shows a possibility of use of empiric Miller formula [4] to determine the MAPD operating parameters. The following modified Miller formula that takes into account voltage drop on serial resistor $R_o$ was used to obtain better agreement of theoretical avalanche gain M dependence on applied bias voltage $U_d$ with experimental data [5]:

$$M = \frac{1}{1-\left(\frac{U_{pn}}{U_{br}}\right)^k} \quad (1)$$

where $U_{br}$ is a breakdown voltage of an APD p-n-junction, $U_{pn} = U_d - J_t \cdot R_0$ – is a voltage drop on the p-n-junction itself, $J_t$ is total APD current, $R_0$ is an effective resistance connected to the APD in series, $k$ – is an empirical parameter that depends on an APD structure and on wavelength of registered light.

## 2. Formulae derivation

Total current $J_t$ of an MAPD under operating conditions can be expressed as

$$J_t = J_s + M \cdot (I_d + I_{ph}) \quad (2)$$

---

where $J_s$ is a surface leakage current that is not being amplified by an avalanche, $I_d$ is a volume dark current that initiates an avalanche, $I_{ph}$ – a photocurrent that initiates an avalanche as well. A value of $I_{ph}$ should be measured at low applied to an MAPD voltages (usually $U_d \sim U_{br}/5$) then $M = 1$ (e.g. the photocurrent value does not depend on the applied voltage). The light wavelength should be chosen such that the light is completely absorbed within the depletion layer of the *p-n*-junction.

An amplified photocurrent $J_{ph}$ when $I_{ph} \gg (J_s + I_d)$ can be expressed as

$$J_{ph} = M \cdot I_{ph} \tag{3}$$

Equation (1) shows that at $M \gg 1$ (for example $M > 10$) value of $U_{pn}$ is close enough to $U_{br}$ which satisfies condition $(\Delta U/U_{br}) \ll 1$ where $\Delta U = U_{br} - U_{pn}$. In this case expression $(U_{pn}/U_{br})^k$ in the denominator of equation (1) can be decomposed into series in small value $(\Delta U/U_{br})$. As a result, leaving the first term only, one obtains a simplified Miler formula:

$$M = \frac{1}{k} \cdot \frac{1}{1 - \frac{U_d - M \cdot I_{ph} \cdot R_0}{U_{br}}} \tag{4}$$

Transforming expression (4) one gets an equation describing photocurrent gain $M$:

$$I_{ph} R_0 \cdot M^2 - (U_d - U_{br}) \cdot M - \frac{U_{br}}{k} = 0 \tag{5}$$

Solving equation (5) one obtains following expressions for the gain and avalanche photocurrent in an MAPD:

$$M = \sqrt{\left(\frac{U_{br} - U_d}{2 I_{ph} R_0}\right)^2 - \left(\frac{U_{br}}{k I_{ph} R_0}\right)} - \frac{U_{br} - U_d}{2 I_{ph} R_0} \tag{6}$$

$$J_{ph} = I_{ph} \cdot \sqrt{\left(\frac{U_{br} - U_d}{2 I_{ph} R_0}\right)^2 - \left(\frac{U_{br}}{k I_{ph} R_0}\right)} - I_{ph} \cdot \frac{U_{br} - U_d}{2 I_{ph} R_0} \tag{7}$$

Unknown parameters $U_{br}$ and $k$ can be determined from the voltage-current dependence of the device at known values of $U_d$ and $I_{ph}$. However, expressions (6) and (7) are rather complicated for their comparison with experimental data. Therefore we will consider two most interesting operation modes of an MAPD.

In the case when $U_d < U_{br}$ the voltage drop on the serial resistor $R_0$ can be neglected. Then using (3) and (4) one will obtain

$$\frac{1}{J_{ph}} = \frac{k}{I_{ph} U_{br}} \cdot (U_{br} - U_d) \tag{8}$$

Expression (8) allows one to determine parameters $U_{br}$ and $k$ using linear extrapolation of the experimental dependence of $(1/J_{ph})$ on applied voltage $U_d$.

In the case when $U_d > U_{br}$ a value of the voltage drop on the serial resistor $R_0$ cannot be neglected because high gain causes significant current through the resistor. The third term of equation (5) can be neglected in this case. This leads to linear dependence of $M$ and $J_{ph}$ on the voltage applied to the photodiode.

$$M = \frac{(U_d - U_{br})}{I_{ph} R_0} \qquad (9)$$

$$J_{ph} = \frac{(U_d - U_{br})}{R_0} \qquad (10)$$

On the other hand, expressions (8-10) can be obtained from a more accurate formula for the gain *M* which is a solution of an avalanche photodiode current continuity equation [5, p.106]:

$$M = \frac{1}{1 - I_{int}} \qquad (11)$$

where $I_{int} = \int_0^W \alpha \times \exp\left[-\int_x^W (\alpha - \beta) dy\right] dx$ is an ionization integral, *W* is thickness of a depleted layer of the *p-n*-junction, *α* and *β* are ionization coefficients for electrons and holes respectively. The breakdown voltage is defined from a condition $I_{int}(U_{pn}=U_{br}) = 1$.

It is obvious that value $I_{int}$ depends on applied voltage $U_d$ and therefore can be decomposed into series on small parameter $z = ((U_{br} - U_{pn}) / U_{br}) << 1$ in other words one can write $I_{int}(z)=1+a·z+…$, where *a* is a derivative of $I_{int}$ at breakdown voltage *($U_{pn}=U_{br}$)*. Taking into account the first term only one will get:

$$M = \frac{1}{a} \cdot \frac{1}{1 - \frac{U_{pn}}{U_{br}}} \qquad (12)$$

where $U_{pn} = U_d - J_t \cdot R_0 \approx U_d - J_{ph} \cdot R_0$ is a value of voltage on the p-n-junction of the device. If $a = k$ then expression (12) is completely equivalent to the expression (4). This means that simplified Miller formula (4) while being empiric has a sufficient physics meaning. Parameter *k* (or *a*) is defined as a rate of change of the ionization integral $I_{int}$.

### 3. Comparison with experimental data

Three samples of avalanche photodiodes from different manufacturers were chosen for an experimental examination of the presented technique: micropixel avalanche photodiode *S10362-11-025U* from Hamamatsu (Japan), avalanche photodiode with a single pixel manufactured by ITC-irst collaboration (Italy), and avalanche photodiode APD-33-128V manufactured by our group in collaboration with Zecotek (Singapore). The *S10362-11-025U* device had *1mm×1mm* active area with a quenching microresistor $R_0 = 220\ k\Omega$ and *1600* pixels/$mm^2$ (pixel dimensions were *25 µm×25 µm*). Device from ITC-irst had a single *40 µm×40 µm* pixel with a quenching microresistor $R_0 = 350\ k\Omega$ in series. The APD-33-128 was a convenient avalanche photodiode with *3mm×3mm* active area.

Total dark current $J_t$ of the *S10362-11-025U* avalanche photodiode measured at $U_d$=15 V did not exceed *30 pA*. A semiconductor light emitting diode with wavelength of about *450 nm* was used as a constant light source. A value of photocurrent that generated an avalanche was $I_{ph}$=120 pA at $U_d$=15 V. Dependence of the photocurrent $J_{ph}$ on applied voltage was measured at the voltage values of $U_d$ >15 V.

Figure 1 shows the inverted photocurrent ($1/J_{ph}$) as a function of applied voltage for the *S10362-11-025U* avalanche photodiode (curve 1). The following values of parameters $U_{br} = 68.78$ V and *k = 1.64* were determined using extrapolation of the linear part of the experimental curve and expression (8). These values were used to plot the same dependence (curve 2) based on expression

(7). An effective resistance $R_0$ was taken equal to the pixel serial resistance $R_p = 220\ \kappa\Omega$. The value of $R_p$ was determined from a direct voltage-current dependence of the device. In addition it was established that increase of the resistance $R_0$ up to $3\ M\Omega$ did not lead to significant change in behavior of the curves in a voltage range $U_d < U_{br}$.

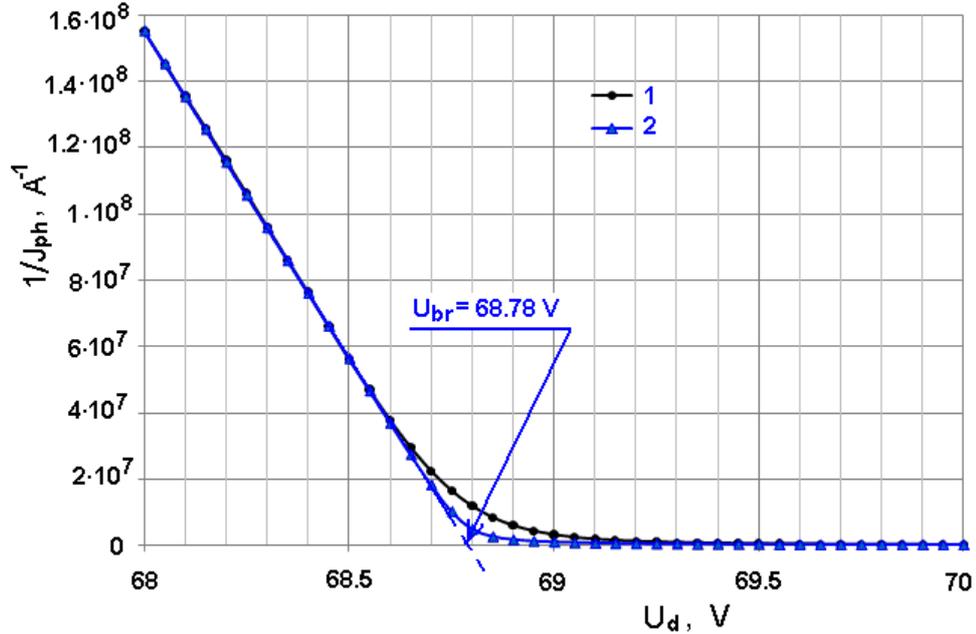

*Figure 1. Dependence of an inverted value of photocurrent on applied voltage. 1 – experimental data, 2 – calculation using Miller formula.*

As one can see from Figure 1 there is good agreement between the calculated and experimental curves when $U_d < U_{br}$. Photocurrent and gain of the avalanche photodiode are described by the Miller formula with sufficient precision in this range of voltages.

Figure 2 shows that when $U_d > U_{br}$ there is significant difference between the results of calculation and experimental data. In this case the Miller formula gives linear dependence of the photocurrent on applied voltage $U_d$ while the experimental curve has close to parabola shape. This is caused by design and operation principles of MAPD. It is well known that an MAPD consists of an array of independent p-n-junctions (pixels) and each pixel is connected to a common bus via individual microresistor $R_p$. Though entire sensitive area is illuminated by constant light each pixel operates in a pulse mode amplifying single photoelectrons. Besides that, the number of fired pixels at constant illumination increases proportionally to $(U_d - U_{br})$. Because of this an effective resistance of the device circuit decreases with increase of $(U_d - U_{br})$ value. Expressing dependence of $(1/R_0)$ value on applied voltage as power series $A + B \times (U_d - U_{br})$ and substituting it into (10) one can get

$$J_{ph} = \frac{(U_d - U_{br})}{R_0} = A \times (U_d - U_{br}) + B \times (U_d - U_{br})^2 \sim (U_d - U_{br})^\delta \qquad (13)$$

where $\delta > 1$ is a parameter, $A$ and $B$ are decomposition coefficients. If the experimental data from the top part of Figure 2 are recalculated as a function $\sqrt[\delta]{J_{ph}} \sim U_d$ then there should be an area of linear dependence. Extrapolation of the linear dependence should cross the abscissa axis at the value of $U_{br}$.

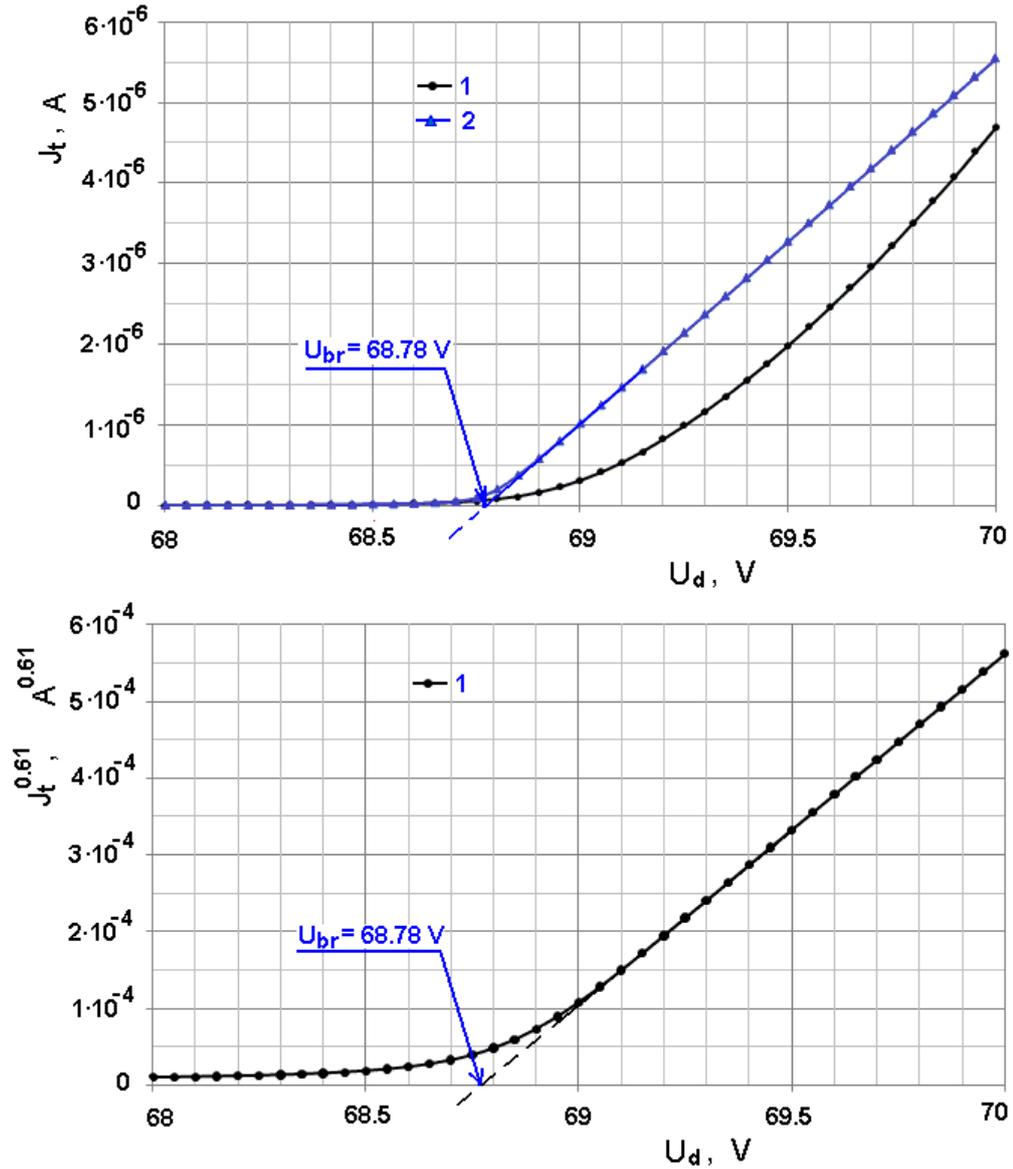

*Figure 2. Dependence of the photocurrent on applied voltage. 1 – experimental data, 2 – calculation using Miller formula.*

Figure 2 (bottom) shows dependence of $\sqrt[\delta]{J_{ph}}$ on $U_d$ when $\delta = 1.64$. Extrapolation of the linear part of the curve crosses a voltage axis in the point of breakdown voltage which can serve as a validation of proposed technique that allows to determine MAPD breakdown voltage.

Further studies were conducted with a single pixel avalanche photodiode with a microresistor in series from ITC-irst [6]. Figure 3 (curve 1) shows dependence of gain *M* on applied voltage $U_d$ (taken from [6]). The device was operating in Geiger mode ($U_d > U_{br} = 30.85$ V) and was registering single photoelectrons. If value of $q/R_0 \cdot C_p = 8.47$ *pA* is substituted as a mean photocurrent $I_{ph}$ into derived from the Miller formula expression (9) the resulting equation gives good agreement with the experimental dependence of *M* on $U_d$ (curve 2). Here *q* is a charge of an electron, $R_0 = 350$ *k*Ω is serial resistance, $C_p = q \times \frac{\partial M}{\partial V_d} = 54$ *fF* – is effective capacitance of the photodiode determined from a slope of experimental dependence of gain *M* on $U_d$.

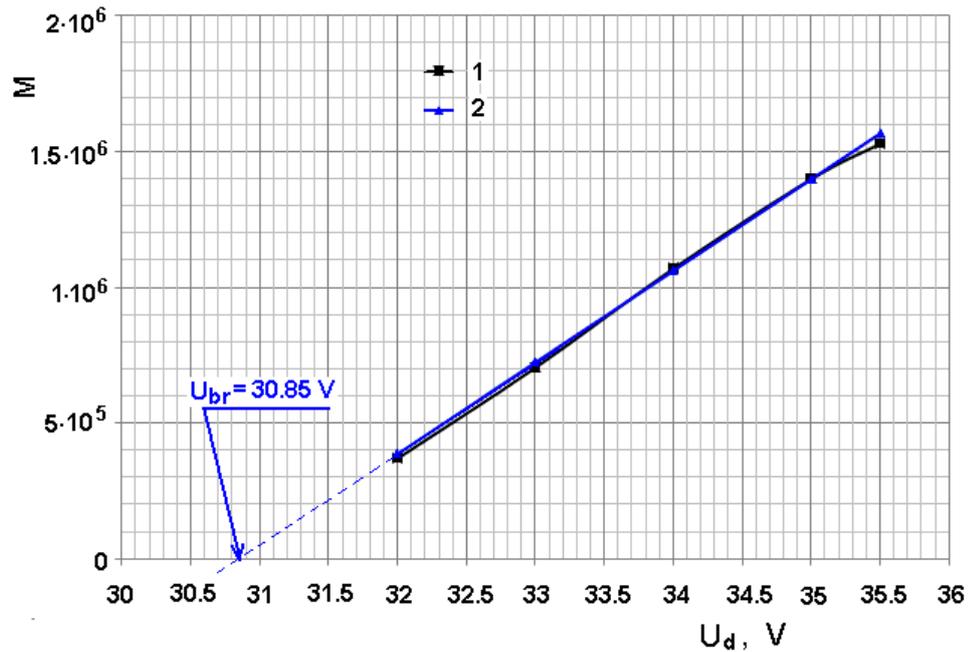

*Рис.3. Dependence of gain of the single pixel photodiode on applied voltage. 1 – experimental data, 2 – calculation using Miller formula.*

Finally, one more device used for verification of the described above technique was avalanche photodiode *APD-33-128* what operated in regular, not Geiger, mode [7]. Figure 4 shows dependence of inverted photocurrent in the device on applied voltage. A value of constant photocurrent able to initiate an avalanche was measured at $U_d = 30\ V$ and was 0.2 nA. The breakdown voltage of the device $U_{br}=129.1\ V$ and Miller parameter $k=0.81$ were determined using extrapolation of curve 1. These values were used in expression (7) with a load resistance of $R_0=1\ k\Omega$. Figure 4 shows that resulting dependence (curve 2) agrees well with the experimental points.

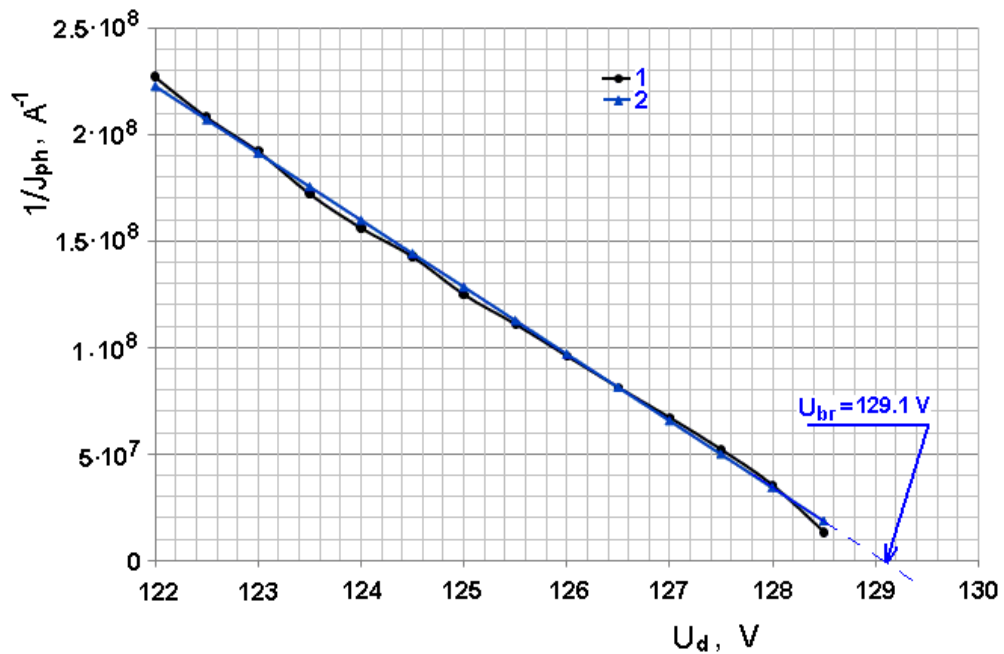

*Рис.4. Inverted photocurrent of the APD-33-128V photodiode as a function of applied voltage. 1 – experimental data, 2 – calculation using Miller formula.*

## 4. Conclusions

Presented calculations and experimental results show that modified Miller formula can be used to determine parameters of avalanche photodiodes.

- Modified Miller formula that takes into account voltage drop on serial resistance describes operation of both regular and micropixel avalanche photodiodes with sufficient accuracy. The Miller parameter must be determined from dependence of photocurrent on applied voltage.
- The breakdown voltage of both regular and micropixel avalanche photodiodes can be determined applying corresponding transformations derived above and using linear extrapolation of obtained curves.